\documentclass{PoS}

\title{Weak low-energy couplings from topological zero-mode wavefunctions}

\ShortTitle{Weak low-energy couplings}

\author{\speaker{Pilar Hern\'andez}\\
        Universidad de Valencia and IFIC-CSIC\\
        E-mail: \email{pilar.hernandez@ific.uv.es}}

\author{Mikko Laine\\
        University of Bielefeld\\
        E-mail: \email{laine@physik.uni-bielefeld.de}}

\author{Carlos Pena\\
        Universidad Aut\'onoma de Madrid and IFT-CSIC\\
        E-mail: \email{carlos.pena@uam.es}}
        
        \author{Emma Torr\'o\\
        Universidad de Valencia and IFIC-CSIC\\
        E-mail: \email{emma.torro@ific.uv.es}}
        
        \author{Jan Wennekers\\
        School of Physics, University of Edinburgh\\
        E-mail: \email{jwenneke@ph.ed.ac.uk}}
        
        \author{Hartmut Wittig\\
        University of Mainz\\
        E-mail: \email{wittig@kph.uni-mainz.de}}
        
\abstract{We discuss a new method to determine the low-energy couplings of the $\Delta S=1$ weak Hamiltonian  in the $\epsilon$-regime. It relies on a matching of the topological poles in $1/m^2$ of three-point functions of two pseudoscalar densities and a four-fermion operator computed in lattice QCD, to the same observables in the Chiral Effective Theory. We present the results of a NLO computation in chiral perturbation theory of these correlation functions together with some preliminary numerical results. }

\FullConference{The XXV International Symposium on Lattice Field Theory\\
		 July 30-4 August 2007\\
		 Regensburg, Germany}

\begin{document}

\section{Introduction}

Non-leptonic kaon decays can be described in the context of a Chiral Effective Theory (ChPT), where the non-perturbative dynamics is encoded in the low-energy 
couplings associated with the $\Delta S=1$ Hamiltonian. These low-energy couplings can  be determined  from first principles in lattice QCD, by performing a matching of suitably chosen 
correlation functions computed in lattice simulations and in the effective theory \cite{oldies}. Such matching should be carried out as close as possible to the chiral limit. 

In \cite{strat}, a new strategy to reveal the role of the charm quark mass in the $\Delta I=1/2$ rule was proposed. The idea is to consider   the ${\rm SU}(4)$-flavour limit, that is a theory with four light quarks corresponding to the $u,d,s$ and $c$, and compute the low-energy couplings of the corresponding $\Delta S=1$ Chiral Hamiltonian.  In a second stage, the charm quark mass is increased towards its physical value, monitoring  the change of the  low-energy couplings. In \cite{prl}, the first determination of the leading-order low-energy couplings of the ${\rm SU}(4)$  $\Delta S=1$ Hamiltonian, $g^\pm$,  was presented. For details on this computation and the precise definition of these couplings we refer
the reader to references \cite{strat, prl}.

In this work we present a new method to determine $g^\pm$ in the so-called $\epsilon$-regime \cite{epsilon,lh}, from correlation functions involving pseudoscalar densities that contain topological poles in $1/(m V)^n$, when evaluated in sectors of fixed, and non-zero, topological charge. 

\section{$g_\pm$  from zero-mode wavefunctions}

In the $\epsilon$-regime and in a fixed topological sector, correlation functions involving quark propagators may contain poles in $1/(mV) ^n$, where $n$ is some integer number, whenever the contribution of the zero-modes to the spectral representation of the quark propagator gives a non-vanishing contribution to the correlation function.  The residues of these poles are easier to compute than the correlation functions themselves. The idea, first explored in \cite{zero_pp}, is then to use the residues of the topological poles to perform the matching, instead of the full correlation function. Given a correlation function $C_\nu(x_1,x_2,...)$, the residue can be isolated by
\begin{eqnarray}
C_{\nu}(x_1,x_2,...) = {Res_n\over (m V)^n} + ... , \;\;\;\;\;\;\;\;Res_n = \lim_{m\rightarrow 0} (m V)^n C_{\nu}(x_1,x_2,...). 
\end{eqnarray}

In \cite{zero_pp} the two-point function of the pseudoscalar density was considered in this context. 
The presence of a pole in $1/(m V)^2$ implies that the corresponding residue can be computed 
fully in terms of the zero-mode wavefunctions, no propagator computation is required. On the effective theory side, the same pole does appear and the residue is a function of only the pseudoscalar decay constant, $F$,  up to NLO.  A numerical exploratory study in the quenched approximation was presented and the usefulness of the method to extract the low-energy coupling $F$ was confirmed. 

In the present work, we extend this idea to the computation of three-point functions from which the weak 
low-energy couplings $g_\pm$ can be extracted. In particular we have considered the following ratios:
\begin{eqnarray}
{ R}_\nu^\sigma &\equiv & { \lim_{m\rightarrow 0} (mV)^2 \sum_{\mathrm{x},\mathrm{y}}\langle \partial_{x_0} P^a(x) O^\sigma(z)
\partial_{y_0} P^b(y)\rangle_\nu \over \lim_{m\rightarrow 0} (mV) \sum_{\mathrm{x}} \langle \partial_{x_0} P^a(x)
{J_L}_0^a(z) ~\rangle_\nu \lim_{m\rightarrow 0} (mV) \sum_{\mathrm{y}} \langle \partial_{y_0} P^b(y)
{J_L}_0^b(z) \rangle_\nu } \nonumber \\
 &\equiv& {{{A}}_\nu(x_0-z_0,y_0-z_0) +\sigma \tilde{{A}}_\nu(x_0-z_0,y_0-z_0) \over
{B}_\nu(x_0-z_0) {B}_\nu(y_0-z_0)}   \quad\sigma = \pm, 
\label{rnu}
\end{eqnarray}
where $P^a \equiv i \bar{\Psi} \gamma_5 T^a \Psi$, ${J_L}_0^a \equiv \bar{\Psi} \gamma_0 P_L T^a \Psi$ and $O^\pm$ are the four fermion operators transforming in the 84 and 20 representation of ${\rm SU}(4)$, for further details see \cite{strat}.
If $v_i(x)$ are the zero-mode wavefunctions of negative chirality, we find:
\begin{eqnarray}
 {A}_\nu(x_0-z_0,y_0-z_0) &\equiv &  {1 \over L^3}~\sum_{\mathrm{x},\mathrm{y},\mathrm{z}}~ \Bigl\langle 
 \sum_{i \in {\cal K}} \eta^\dagger_i(x) S(x,z) \gamma_\mu P_- v_i(z) 
 \sum_{j \in {\cal K}} \eta^\dagger_j(y) S(y,z) \gamma_\mu P_- v_j(z) 
 \Bigr\rangle_\nu 
\nonumber  \\ 
 \tilde{A}_\nu(x_0-z_0,y_0-z_0) & \equiv & -{1 \over L^3}~\sum_{\mathrm{x},\mathrm{y},\mathrm{z}} \Bigl\langle 
 \sum_{i,j \in {\cal K}} \eta^\dagger_j(x) S(x,z) \gamma_\mu P_- v_i(z) \eta^\dagger_i(y) S(y,z) \gamma_\mu P_- v_j(z) 
 \Bigr\rangle_\nu ,
 \label{aatilde}
\end{eqnarray}
where $\eta_i^\dagger(x) S(x,z) \equiv {v_i^\dagger(x+a \hat{0}) S(x+a\hat{0},z)  - v_i^\dagger(x -a \hat{0}) S(x-a \hat{0},z)  \over 2 a}$ and $S(x,y)$ is the quark propagator  \footnote{Similar expressions are obtained for the opposite chirality.}. It is clear from eqs.~(\ref{aatilde}) that a number of inversions equal to twice the topological charge, i.e. $2 |\nu|$, (with sources $\eta_i(x)$ and $\eta_i(y)$, since $x_0$ and $y_0$ need to  be fixed) is sufficient to construct the correlation function, whilst averaging over all the spatial positions of the three sources. Such averaging was only possible in the standard method of \cite{strat} through low-mode averaging (LMA), and only for the contribution of the low-modes. The price of  LMA is $12+2\times N_{low}$ inversions, where $N_{low}$ was the number of low modes treated separately. Typically $N_{low}$ can be as large as 20, and hence the numerical cost can be quite substantial.
 
 The matching of the  amplitudes in eqs.~(\ref{aatilde}) to the Chiral Effective Theory results in the following relation
 \begin{eqnarray}
g_{\pm} {{\mathcal R}}_\nu^\pm \simeq \left[k^\pm(M_W)\right]_{RGI}~\left[{Z^\pm \over Z_A^2}\right]_{RGI} { R}_\nu^\pm ,
\end{eqnarray}
where $k^\pm(M_W)$ are the Wilson coefficients of the $O^\pm$ operators, $Z^\pm$ are their corresponding renormalization factors and $Z_A$ is the renormalization factor of the axial current. The RGI values of these factors have been computed non-perturbatively in \cite{renorm}. 
On the left-hand side of the equation, ${ {\mathcal R}}^\pm_\nu$ is the ChPT prediction  
for the ratios of eq.~(\ref{rnu}) in the $\epsilon$ regime. At NLO, ${ {\mathcal R}}^\pm_\nu$  only depend on $F$ and the volume, as we show in the next section.

\section{Prediction in Chiral Perturbation Theory at NLO}

\begin{figure}
\begin{center}
\includegraphics[width=.4\textwidth]{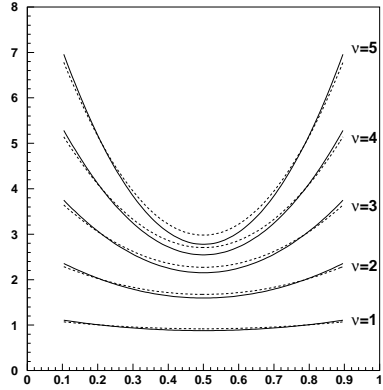}
\caption{$T {\mathcal B}_\nu(x_0)$ for $N=2$ (solid) and $N=0$ (dashed) as a function of $x_0/T$. In both cases we take $F=93$~MeV and $L=2$ fm. }
\label{fig:jp}
\end{center}
\end{figure}

The result of the NLO computation of the two-point function in the denominator of eq.~(\ref{rnu}) in the $\epsilon$ expansion is 
\begin{eqnarray}
T {\mathcal B_\nu}^{(q)}(x_0) &= & |\nu| \left[ 1-  {1\over 12} \rho \frac{|\nu| \left(+ {1 \over N}\right)}{ (F L)^2}  + \rho \frac{|\nu| \left(+{1 \over N}\right)}{ (F L)^2} \left({|x_0| \over T} 
- {1\over 2}\right)^2 \right],
\label{bnuchpt}
\end{eqnarray}
where $\tau \equiv x_0/T$, $\rho \equiv T/L$  and $N$ is the number of dynamical flavours. The quenched result,
${\mathcal B}^{q}_\nu$, is the same leaving out the term in parenthesis $\left(+{1\over N}\right)$. The first term corresponds to the LO result, that is the constant $|\nu|$. In Figure~\ref{fig:jp} we show the result for this quantity at NLO  for different values of $|\nu|$ in a symmetric box of size $L=2$~fm. 

The results for the ratios  of three to two-point functions are
\begin{eqnarray}
{{\mathcal R}^\pm}^{(q)}_\nu(x_0-z_0,y_0-z_0) &\equiv &  \left(1 \mp \frac{1 }{|\nu|}\right) \left[ 1 \pm 2 { \beta_1\over F^2 \sqrt{V}} \right.\nonumber\\
& &\left. \pm {T^2\over F^2  V} \left( g_1(\tau_x)+ g_1(\tau_y) -h_1(\tau_x) -h_1(\tau_y) + \left(1 \left(\mp {2 \over N}\right)\right) H(\tau_x,\tau_y) \right)\right],\nonumber\\
\end{eqnarray}
where $\tau_x= (x_0-z_0)/T$, $\tau_y= (y_0-z_0)/T$, $\beta_1$ is a constant \cite{epsilon},  and 
\begin{eqnarray}
h_1(\tau) &\equiv & {1 \over 2} \left( \left(|\tau|-{1\over 2}\right)^2 -{1 \over 12} \right),\\
g_1(\tau) &=&   \left[h'_1(\tau)\right]^2 + \sum_{\vec{n} \neq 0} \left[{ A'_p(\tau)^2 +|\vec{p}|^2 A_p(\tau)^2}\right],\\
H(\tau_x,\tau_y) &\equiv & h_1'(\tau_x) h_1'(\tau_y)   -h_1(\tau_x-\tau_y)  - (h_1'(\tau_x) -h_1'(\tau_y)) h_1'(\tau_x-\tau_y) ,
\end{eqnarray}
where $|\vec{p}| = 2 \pi \rho |\vec{n}|$ with $\vec{n} = (n_1, n_2, n_3)$ a vector of natural numbers, the primes refer to derivatives with respect to $\tau$, and 
\begin{eqnarray}
A_p(\tau) \equiv {\cosh\left( |\vec{p}| \left(|\tau|-{1 \over 2}\right)\right)  \over 2 |\vec{p}| \sinh(|\vec{p}|/2)}.
\end{eqnarray}
Again the expressions  in the quenched theory are identical at this order, except that the term $\left(\mp {2 \over N}\right)$ is not there. 

The ratios are therefore constants at LO, while at NLO they depend on the insertion of the sources. There is also a significant dependence on $|\nu|$ already at LO. Both these features of the chiral corrections
in ${ {\mathcal R}}^\pm_\nu$ differ from those of the ratios constructed out of  left-current correlators, used in \cite{strat}, where no 
$|\nu|$ dependence, nor  temporal dependence was found at NLO. The large difference between the
chiral corrections in both cases probably implies different systematic uncertainties. Different systematics
in the two approaches serve as additional consistency checks of the method. 

\begin{figure}
\begin{center}
\includegraphics[width=.4\textwidth]{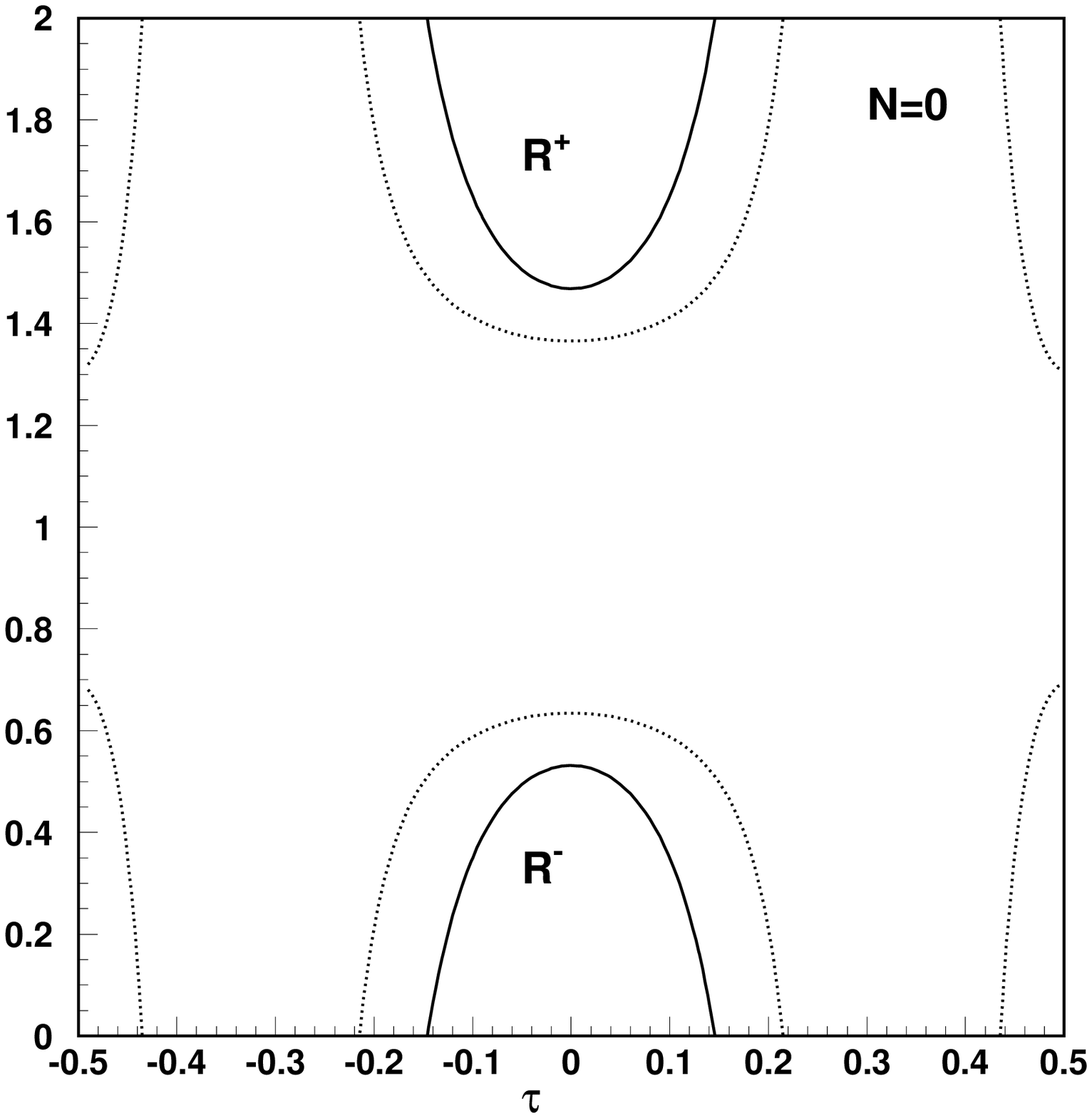}\includegraphics[width=.4\textwidth]{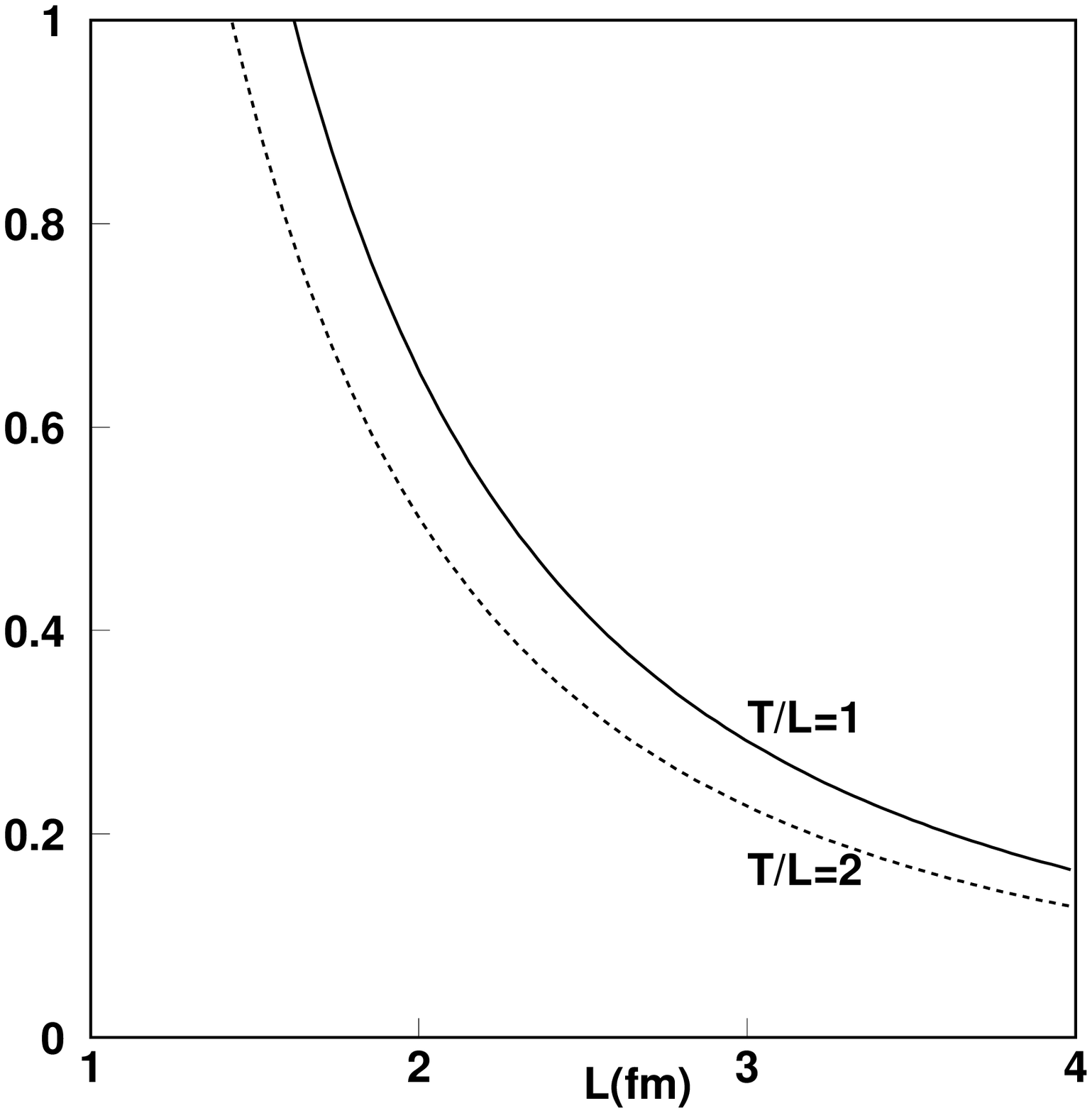}
\caption{Left: ${{\mathcal R}^\pm}^{q}_\nu(T/3-\tau T,2 T/3 -\tau T)/\left(1\mp {1 \over |\nu|}\right)$ for $N=0$ and $L=2$ fm with $T/L=1$ (solid) and $T/L=2$ (dashed) as a function of $\tau \equiv z_0/T$.  Right: $\left|{{\mathcal R}^\pm}^{q}_\nu(T/3,2 T/3)/\left(1\mp {1 \over |\nu|}\right)-1\right|$ as a function of $L$.}
\label{figure:rpm}
\end{center}
\end{figure}

In the left plot of Figure~\ref{figure:rpm} we show the result for ${{\mathcal R}^\pm}^{q}_\nu(T/3 -\tau T,2T/3 -\tau T)/(1\mp {1\over |\nu|})$ in a box of $L=2$~fm for $\rho=1$ and $\rho=2$.
Unfortunately NLO corrections seem to be rather large still at $2$ fm, as shown on the right plot of Figure~\ref{figure:rpm}. 

\section{Exploratory quenched study}

We have recently carried out an exploratory study of the two and three-point functions in eq.~(\ref{rnu}) in the quenched approximation. We have considered the simulation parameters of Table~\ref{tab:param}, which correspond to two lattices with the same physical volume of around (2 fm)$^4$, and different lattice spacings.

We have used the overlap operator. For all the details on the implementation and algorithms we refer the reader to related previous work \cite{num,zero_pp,strat}. 
Here we will only present our preliminary results for the two-point functions. The results for the three-point functions will be presented in detail in a forthcoming publication. 

\begin{figure}
\begin{center}
\includegraphics[width=.4\textwidth]{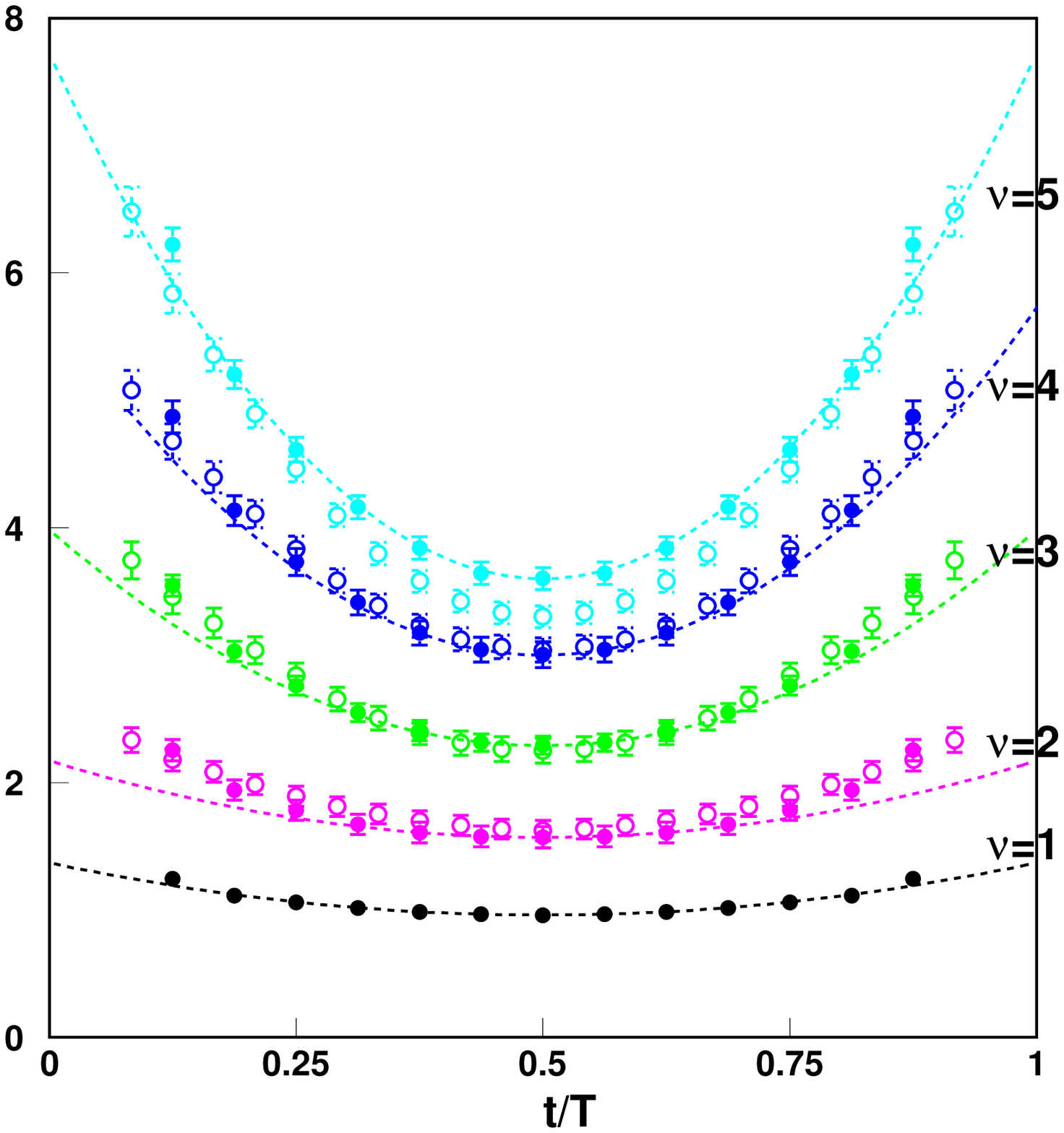}\includegraphics[width=.4\textwidth]{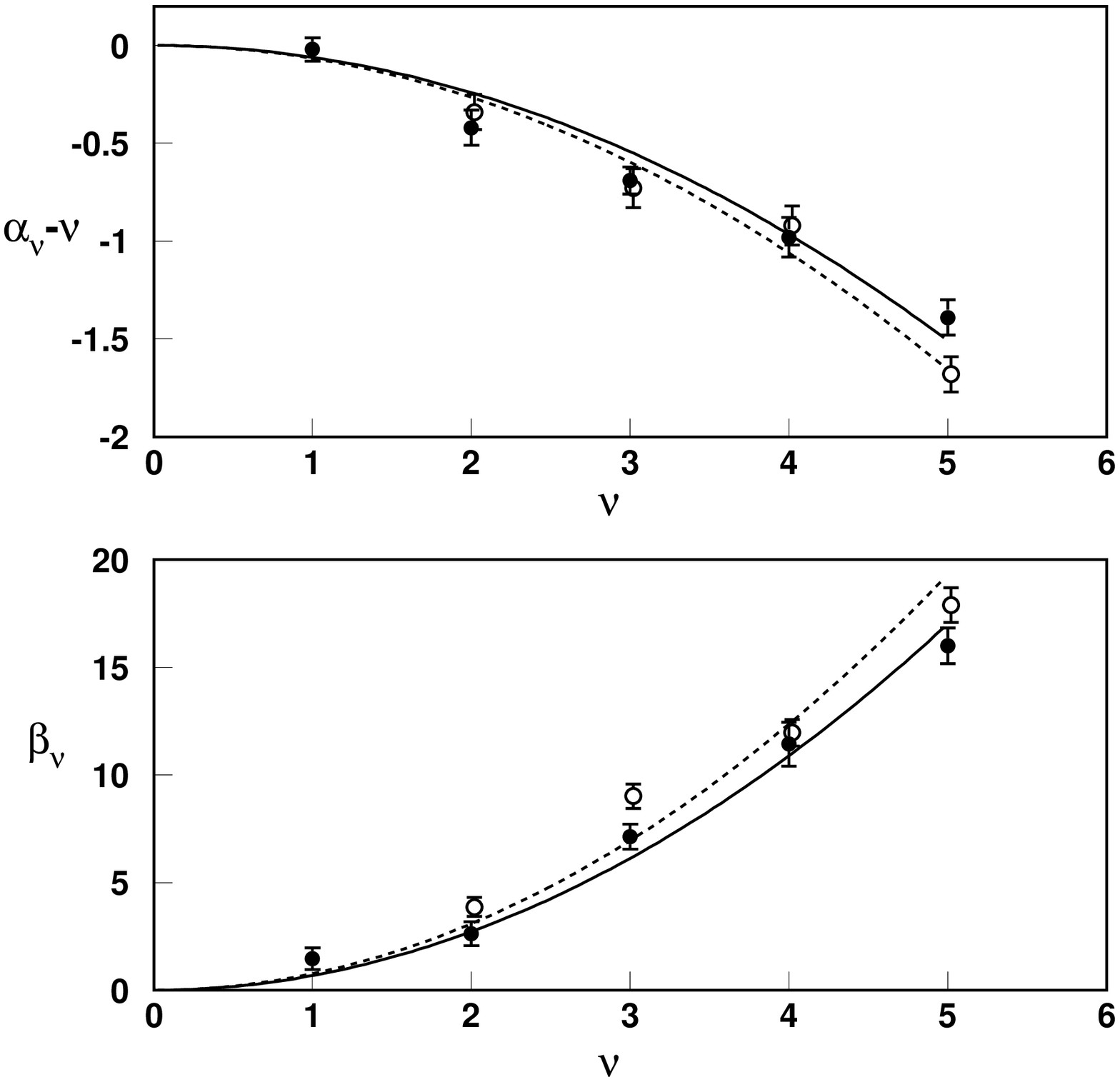}
\caption{Left: $T B_\nu(t)$ as a function of $t/T$ for the lattice A1 (full symbols) and B1 (empty symbols). 
The curves are fits of the A1 data to NLO ChPT. Right: $\alpha_\nu-|\nu|$ and $\beta_\nu$ as a function of $|\nu|$ for lattice A1 (full symbols) and B1 (empty symbols). The curves are fits to the NLO expectation for A1 (solid) and B1(dashed). }
\label{fig:2pt}
\end{center}
\end{figure}

 In the left plot of Figure~\ref{fig:2pt}, we show the results for $B_\nu(t)$ as a function of $\tau=t/T$ compared with a fit of the form 
 \begin{eqnarray}
 {\mathcal B}_\nu(t) = \alpha_\nu + \beta_\nu \left({t \over T} - {1 \over 2}\right)^2, \label{eq:fitform}
\end{eqnarray}
where the NLO ChPT predictions of eq.(\ref{bnuchpt}) imply
\begin{eqnarray}
\alpha_\nu = |\nu| -{1\over 12} {T\over L} {|\nu|^2 \over (FL)^2},~~~ \beta_\nu= {T\over L} {|\nu|^2 \over (FL)^2} .
\label{eq:alphabeta}
\end{eqnarray}
We show in the right plot of Figure~\ref{fig:2pt}, the dependence of $\alpha_\nu-|\nu|$ and $\beta_\nu$ on $|\nu|$, together with the fit to a parabola from which $F L$ can be extracted. Although not very precise, the value of $F$ so obtained is in agreement with more direct determinations. Discretization effects in these observables are quite small.
\begin{table}
\begin{center}
\begin{tabular}{ccccccc}
\hline\\[-2.0ex]
Lattice & $\beta$ & $V$ & $|\nu|$ & $N_{conf}$ & $x_0/a,y_0/a$\\[1.5ex]
\hline\\[-2.0ex]
A1 & 5.8458 & $16^4$ & 1--5 &  282 & 5,11\\
B1 & 6.0375 & $24^4$ & 2--5 &  236 & 8,16 \\
\hline
\end{tabular}
\caption{Simulation parameters}
\label{tab:param}
\end{center}
\end{table}
 \begin{figure}
\begin{center}\includegraphics[width=.4\textwidth]{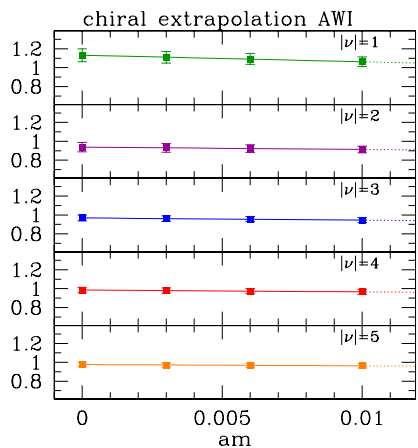}
\vspace*{-1cm}
\caption{Check of the truncated Ward Identity.}
\label{fig:wi}
\end{center}
\end{figure}
 An interesting check of the two-point functions is provided by the chiral Ward identity, which implies the relation 
 \begin{eqnarray}
Z_A {B}_\nu(x_0-y_0) =  \lim_{m\rightarrow 0} m^2 V  \sum_{\vec{x}}~ \langle{P}^a(x)~{P}^a(y)\rangle_\nu .
\label{eq:wi}
\end{eqnarray}
The right-hand side of  eq.~(\ref{eq:wi}) is the topological-pole contribution of the pseudoscalar two-point function, that is the observable that was studied in \cite{zero_pp}. Since $Z_A$ has been computed before \cite{ww}, we can compare the two sides of the equation.
The ratio of the left over the right-hand side of the equation is shown in Figure~\ref{fig:wi} for the coarser 
lattice. The result  is one in the chiral limit, as expected, in all topological sectors studied. Although it is no 
surprise that the Ward identity is satisfied given the exact chiral symmetry of the discretization, it is 
a non-trivial test of the method that it is also satisfied when both sides of the identity are truncated 
to the zero-mode pole contributions, especially because the limit $m\rightarrow 0$ can be done analytically on the right-hand side but not on the left-hand side. 
 
 
 
\acknowledgments{We wish to thank L.~Giusti, M.~L\"uscher and P. Weisz for allowing us to use part of the code developed jointly. 
We acknowledge the computer resources provided by IBM MareNostrum at the BSC, the IBM Regatta at FZ J\"ulich and the PC-clusters at 
University of Valencia. P. H. and E. T. acknowledge partial finantial support from the research grants: FPA-2004-00996, FPA-2005-01678, FLAVIAnet and HA2005-0120.}

\end{document}